\documentclass[onecolumn]{emulateapj}

\slugcomment{}

\shorttitle{Gravitational Wave from Supermassive Black Hole  Coalescence}
\shortauthors{Enoki et al.}

\begin{document}

\title{Gravitational Waves from Supermassive Black Hole Coalescence 
in a Hierarchical Galaxy Formation Model}

\author{Motohiro Enoki \altaffilmark{1}, Kaiki T. Inoue
\altaffilmark{1,2}, Masahiro Nagashima \altaffilmark{3,4} and Naoshi
Sugiyama
\altaffilmark{1}}

\email{enoki.motohiro@nao.ac.jp}
\email{kinoue@phys.kindai.ac.jp}
\email{masa@scphys.kyoto-u.ac.jp}
\email{naoshi@th.nao.ac.jp}

\altaffiltext{1}{National Astronomical Observatory, Osawa 2-21-1, Mitaka, Tokyo, 181-8588, Japan}

\altaffiltext{2}{Present address: Department of Natural Science and 
Engineering, 
Kinki University, Higashi-Osaka, Osaka, 577-8502, Japan}

\altaffiltext{3}{Department of Physics, 
University of Durham, South Road, Durham DH1 3LE, U.K.  }

\altaffiltext{4}{Present address: Department of Physics, 
Kyoto University, Sakyo-ku, Kyoto, 606-8502, Japan}

\begin{abstract}
We investigate the expected gravitational wave emission 
from coalescing supermassive
black hole (SMBH) 
binaries resulting from mergers of their host galaxies.
When galaxies merge, the SMBHs in the host galaxies 
sink to the center of the new merged galaxy and form a binary system.
We employ a semi-analytic 
model of galaxy and quasar formation  based on the
hierarchical clustering scenario to estimate the
amplitude of the expected stochastic 
gravitational wave background owing to inspiraling SMBH binaries 
and bursts owing to the SMBH binary coalescence events.
We find that the characteristic
strain amplitude of the background radiation is  
$h_c(f) \sim 10^{-16} (f/1 \mu {\rm Hz})^{-2/3}$ for $f \lesssim 1 \mu {\rm Hz}$ just below the detection limit from measurements of the pulsar timing
provided that SMBHs coalesce simultaneously when host galaxies merge.
The main contribution to the total strain amplitude of the background
 radiation comes from SMBH coalescence events at $0<z<1$.
We also find that a future 
space-based gravitational wave interferometer such as the planned 
\textit{Laser Interferometer Space Antenna} ({\sl LISA}) might
detect intense gravitational wave bursts associated with coalescence of SMBH
binaries with total mass $M_{\rm tot} < 10^7 M_{\odot}$ at $z \gtrsim 2$
 at a rate $ \sim 1.0 \ {\rm yr}^{-1}$. 
Our model predicts that burst signals 
with a larger amplitude $h_{\rm burst} \sim 10^{-15}$
correspond to coalescence events of 
massive SMBH binary with total mass $M_{\rm tot} \sim 10^8
M_{\odot}$ at low redshift  $ z \lesssim 1$ 
at a rate $ \sim 0.1 \ {\rm yr}^{-1}$
whereas those with a smaller amplitude $h_{\rm
 burst} \sim 10^{-17}$ correspond to coalescence events of 
less massive SMBH binary with total mass
$M_{\rm tot} \sim 10^6 M_{\odot}$ 
at high redshift $ z \gtrsim 3$.
\end{abstract}


\section{Introduction}\label{sec:intro}
In recent years, there has been  increasing observational evidence that many nearby galaxies have central supermassive black holes (SMBHs)
in the mass range of $10^6-10^9 M_{\odot}$,
 and that their physical properties
correlate with those of 
spheroids\footnote{Throughout this
 paper, we refer to bulge or elliptical galaxy as {\it spheroid}.} of
their host galaxies. First,  
the estimated mass of a SMBH in a galactic center, $M_{\rm BH}$, is
roughly proportional to the mass
of the spheroid, $M_{\rm spheroid}$. The ratio of $M_{\rm BH}/M_{\rm spheroid}$ is
$0.001-0.006$ in each galaxy
(e.g. \citealp{Kormendy95}; \citealp{Magorrian98};
\citealp{Merritt01b}). Second, the mass of a SMBH  
correlates with the velocity dispersion of stars in the spheroid, $\sigma_{\rm spheroid}$, as $M_{\rm BH} \propto
\sigma_{\rm spheroid}^{n},n=3.7-5.3$ (e.g. \citealp{Ferrarese00}; \citealp{Gebhardt00};
\citealp{Merritt01a}; \citealp{Tremaine02}).
These relations suggest that the
formation of SMBHs physically links to the formation of spheroids that
harbor the SMBHs. 

In order to study the
formation and evolution of SMBHs, it is necessary to construct a model
that includes galaxy formation and merger processes.
It has become widely accepted that quasars are fueled by 
accretion of gas onto SMBHs in the nuclei of host galaxies. 
Hydrodynamical simulations have shown that a merger of galaxies drives
gas fall rapidly onto the center of the merged system 
and fuels a nuclear starburst leading to spheroid formation 
\citep[e.g.][]{Mihos94,Mihos96,Barnes96}.
Some observations of quasar hosts 
show that many quasars reside in spheroids of interacting
systems or elliptical galaxies \citep[e.g.][]{Bahcall97,McLure99,Dunlop03,Kauffmann03}.
Thus the major merger of galaxies is a possible mechanism for quasar
activity, SMBH evolution
and spheroid formation.  

In the standard hierarchical structure formation scenario in a cold dark matter (CDM) universe, dark-matter halos ({\it dark halos}) cluster gravitationally
and merge together. In each of merged dark halos, a galaxy 
is formed as a result of  
radiative gas cooling, star formation, and  
supernova feedback. 
Several galaxies in a common dark halo sometimes merge together and a
more massive galaxy is assembled. When galaxies merge, SMBHs in the
centers of the galaxies sink toward the center of the new merged galaxy and 
form a SMBH binary subsequently. 
If the binary loses enough energy and angular 
momentum, it will evolve to the gravitational wave emitting 
regime and begin inspiraling, eventually coalesces with a 
gravitational wave burst. 

An ensemble of 
gravitational waves from a number of inspiraling 
SMBH binaries at different redshift can be observed
as a stochastic background at frequencies
$\sim 1{\rm n} - 1 \mu {\rm Hz}$, which can be detected by 
pulsar timing measurements \citep{Detweiler79}.
Gravitational wave bursts from coalescence events 
can be detected by a Doppler tracking test of 
interplanetary spacecraft \citep{Thorne76}. Future 
space interferometers such as the Laser Interferometer Space 
Antenna \textit{LISA} might detect quasi-monochromatic 
wave sources associated with inspiraling SMBH binaries 
and gravitational wave bursts associated
with SMBH binary coalescence.

In order to estimate the feasibility of detecting
such gravitational waves, we need to know the SMBH coalescing rate that depends on the
complex physical processes before coalescence.
A possible scenario of forming SMBHs consists of four stages:
First, a seed massive black hole in the center of a galaxy 
grows to an intermediate mass black hole (IMBH)
 via run-away collapse in a dense star cluster
\citep{Zwart04}.  
Second, a galaxy that contains an IMBH in their 
center merges with another galaxy with an IMBH. 
Third, the two IMBHs sink to the center of gravitational potential well
owing to dynamical friction until a ``hardening'' regime \citep{Volonteri03}.
Finally, physical processes such as three-body processes, 
gas dynamics, the ``Kozai'' mechanism, etc. drive the binary
to the gravitational wave emission regime
\citep[e.g.][]{Begelman80,Makino97s,Gould00,Yu02,Armitage02,Blaes02,Volonteri03,Milosavljevic03}.
However , the efficiency to bring two SMBHs together is still unknown. 

To date, a number of attempts have been
made to calculate the SMBH coalescence rate 
taking the physical processes 
in the ``second'' and ``third'' stages into account. 
Some authors use phenomenological models of
galaxy mergers based on number counts of quasars 
and spheroids \citep[e.g.][]{Thorne76,
Fukushige92, Haehnelt94, Rajagopal95, Jaffe03}. Others use 
merger rates of dark halos (not galaxies) based on formalism of \cite{PS74} and \cite{Lacey93}
\citep[e.g.][]{Haehnelt94, Menou01}. 
\cite{Volonteri03} calculated the SMBH coalescence rate taking 
conditions under which ``sub-halos'' with SMBHs in a dark halo sink to
the dark halo center into account \citep[see
also][]{Wyithe03,Sesana04}. 
However, none of these models include baryonic gas evolution 
and galaxy formation processes.
Because SMBH formation process is relevant to spheroids of host
galaxies rather than to dark halos, we need to evaluate 
how the baryonic gas processes such as star formation and radiative
cooling affect the SMBH coalescence rate.

In this paper, we estimate the SMBH coalescence rate 
using a new semi-analytic (SA) model of \cite{Enoki03}
(an extended model of \citet{Nagashima01b})
in which the SMBH formation is incorporated into the galaxy formation. 
Then, we calculate the spectrum of gravitational wave
background from inspiraling SMBH binaries, based on the formulation given by
\cite{Jaffe03} and we compare our result with 
that from a pulsar timing measurement.
We also estimate the event rate of gravitational wave bursts from
SMBH coalescence events that might be detected by future planned space laser 
interferometers, based on an argument in \cite{Thorne76}. 

In SA models, merging histories of dark  halos are realized  using a
Monte-Carlo algorithm  and evolution of baryonic components within
dark halos is calculated using simple analytic models for gas cooling,
star formation, supernova feedback, galaxy merging and other processes. 
SA models have successfully reproduced a variety of observed
 features of galaxies, such as their luminosity functions, color
 distributions, and so on \citep[e.g.][]
{KWG93,Cole94,Cole00,SP99,Nagashima01b,Nagashima02}.
\cite{KH00} introduced a unified model that includes
 the formation of both galaxies and SMBHs within the framework of their
 SA model \citep[see also][]{Cattaneo01,Menci03,Islam03,Granato04}. 
Our SA model reproduces not only these observational features but also  
the observed present black 
hole mass function and the quasar luminosity functions at different
redshifts \citep{Enoki03}. 

The paper is organized as follows: in section \ref{sec:merger}  
we briefly review our SA model for galaxy formation and SMBH growth; in
section \ref{sec:GWR} we calculate the spectrum of gravitational wave
background and the event rate of gravitational wave bursts; in
section \ref{sec:summary} we provide summary and conclusions.

\section{Galaxy Merger / Black Hole Coalescence Rate}\label{sec:merger}
 In this section we briefly describe our SA model for galaxy formation
 and the SMBH growth. The details are shown in
 \cite{Nagashima01b} and \cite{Enoki03}. 
 
\subsection{Model of Galaxy Formation}\label{subsec:galmodel}
First, we construct Monte Carlo realizations of merging histories of
 dark halos from the present to higher redshifts using a method of \citet{SK99}, which is 
 based on the extended 
Press--Schechter formalism \citep{PS74,Bond91,Bower91,Lacey93}. 
Merging histories of dark halos depend on the 
 cosmological model. The adopted cosmological model is a low-density,
 spatially flat cold dark matter ($\Lambda$CDM) universe with
the present density parameter, $\Omega_{\rm m}=0.3$, the cosmological
constant, $\Omega_{\Lambda}=0.7$, the Hubble constant $h=0.7$ ($h \equiv
 H_0/100 \; {\rm km \
 s^{-1}\ {Mpc^{-1}}}$) and the present rms density fluctuation in spheres
 of $8 h^{-1} {\rm Mpc}$ radius, $\sigma_8=0.9$. Dark halos with
circular velocity, $V_{\rm circ}<$40 km~s$^{-1}$, are treated as diffuse
 accretion matter. This condition comes from the estimation of Jeans
 mass in the ultraviolet background radiation field \citep[e.g.][]{Thoul96}. 
The adopted timestep of merging histories of dark halos is a redshift interval of $\Delta z=0.06(1+z)$,
corresponding to dynamical time scale of dark halos which collapse
 at redshift $z$. The highest redshift in each merging path which depends on
 the present dark halo mass, is about $z \sim 20-30$. 

Next, in each merging path of dark halos, we calculate the
evolution of the baryonic component from higher redshifts to the
present. If a dark halo has no progenitor halos, the mass
fraction of the gas in the halo is given by $\Omega_{\rm b}/\Omega_{\rm
m}$ where $\Omega_{\rm b}$ is the baryonic density parameter. Here we use $\Omega_{\rm b} = 0.02 h^{-2}$. When a dark halo
 collapses, the gas in the halo is shock-heated to the virial
 temperature of the halo (the {\it hot gas}). At the same time, the gas
 in dense regions of the halo cools owing to efficient 
radiative cooling and sinks to the center of
 the halo and settle into a rotationally supported disk until the
 subsequent collapse of the dark halo.  We call this cooled
gas the {\it cold gas}. Stars are formed from the cold gas at a rate of $\dot{M}_{*}={M_{\rm
cold}}/{\tau_{*}}$, where $M_{\rm cold}$ is the mass of the cold gas and
$\tau_{*}$ is the time scale of star formation. We assume 
$\tau_{*}=\tau_{*}^{0} (V_{\rm circ}/300  {\rm km~s}^{-1})^{\alpha_{*}} $. 
The free parameters of $\tau_{*}^{0}$ and $\alpha_{*}$ 
are chosen to match the observed mass fraction of cold gas in the disks of spiral galaxies. With star formation, supernovae occur
and heat up the surrounding cold gas to the hot gas phase (supernova
feedback).  The reheating rate is given by  $\dot{M}_{\rm
reheat}=\beta(V_{\rm circ}) \dot{M}_{*}$, where $\beta(V_{\rm circ}) =
(V_{\rm hot}/V_{\rm circ})^{\alpha_{\rm hot}}$. The free parameters of $V_{\rm hot}$ and $\alpha_{\rm hot}$ 
are determined by matching the observed local luminosity function of
galaxies. Given the star formation rate as a function of time, we
calculate the luminosity and color of each galaxy from the
star formation history of the galaxy by using a
stellar population synthesis model. We use the population synthesis code
by \citet{Kodama97}.

When several progenitor halos merge, a newly formed larger
dark halo contains at least two or more galaxies 
which originally resided in the individual progenitor halos.
 We identify the central galaxy in the new common halo with the
central galaxy contained in the most massive progenitor halos. 
Note that cooled hot gas accretes to only central galaxies.
 Other galaxies are regarded as satellite galaxies.
These galaxies merge by either dynamical friction or random collision.
Satellite galaxies merge with the central galaxy owing to dynamical
friction in the following time scale, 
\begin{equation}
\tau_{\rm fric}=\frac{260}{\ln\Lambda_{\rm c}}\left(\frac{R_{\rm H}}{\rm Mpc}\right)^{2}
\left(\frac{V_{\rm circ}}{10^{3}{\rm km~s}^{-1}}\right)
\left(\frac{M_{\rm sat}}{10^{12}M_{\odot}}\right)^{-1}{\rm Gyr},
\end{equation}
where $R_{\rm H}$ and $V_{\rm circ}$ are the radius and the circular
velocity of the new common halo, respectively, $\ln\Lambda_{\rm c}$ is
the Coulomb logarithm, and $M_{\rm sat}$ is the mass of the satellite galaxy
including its dark halo \citep{BT87}.  When the
time  passing after a galaxy became a satellite exceeds $\tau_{\rm
fric}$, the satellite galaxy infalls onto the central galaxy. On the
other hand, satellite galaxies merge with each other in
timescale of random collision. Under the condition that
 the satellite galaxies gravitationally bound and merge during
 encounters, the collision time scale is \citep{Makino97},
\begin{eqnarray}
\tau_{\rm coll}&=&\frac{500}{N^{2}}\left(\frac{R_{\rm H}}{\mbox{Mpc}}
\right)^{3}\left(\frac{r_{\rm gal}}{0.12 \mbox{ Mpc}}\right)^{-2}
\nonumber\\
&&\qquad\times\left(\frac{\sigma_{\rm gal}}{100 \mbox{ km~s}^{-1}}
\right)^{-4}\left(\frac{\sigma_{\rm halo}}{300 \mbox{ km~s}^{-1}}
\right)^{3}\mbox{Gyr},
\end{eqnarray}
where $N$ is the number of satellite galaxies, $r_{\rm gal}$ is a
radius of a satellite, and $\sigma_{\rm halo}$ and $\sigma_{\rm gal}$ are the 1D
velocity dispersions of the common halo and the satellite galaxy,
respectively.
 With a probability of $\Delta t/\tau_{\rm
coll}$, where $\Delta t$ is the timestep corresponding to the redshift
interval $\Delta z$, a  satellite galaxy merges with another randomly
picked satellite.
Let us consider the case that two galaxies of masses $m_1$ and $m_2 (>m_1)$
merge together.  If the mass ratio, $f=m_1/m_2$, is larger than a certain
critical value of $f_{\rm bulge}$ (major merger), we assume that a
starburst occurs: all of the cold gas turns into stars and hot gas 
which fills the dark halo, and
all of the stars populate the bulge of a new galaxy.  On the other hand, if
$f<f_{\rm bulge}$ (minor merger), no starburst occurs and a smaller galaxy is simply
absorbed into the disk of a larger galaxy.  These processes are repeated
until the output redshift.
We classify galaxies into different morphological types according to the
$B$-band bulge-to-disk luminosity ratio, $B/D$.  In this paper, galaxies
with $B/D > 2/3$, and $B/D  \le 2/3$ are classified as ellipticals/S0s and
spirals, respectively. The parameter $f_{\rm bulge}$ is fixed by a 
comparison with the observed type mix.

Model parameters are determined  by a comparison with observations of
galaxies in the local Universe, such as 
luminosity functions and the cold gas mass fraction in spiral galaxies.
Our SA model can reproduce galaxy number counts and
photometric redshift distribution of galaxies in the Hubble Deep Field.
The adopted parameters of
this model are tabulated in table \ref{table1}. Some of the model
parameters ($\sigma_8, \Omega_{\rm b}$ and fraction of invisible stellar
mass $\Upsilon$) are updated and slightly
  different from our previous paper \citep{Nagashima01b,Enoki03}. The
  update of $\sigma_8$ and $\Omega_{\rm b}$ causes only a slight
  change of $\Upsilon$ and we have confirmed
  that the modification hardly affects our results.
Using this SA model, we can estimate 
the galaxy merger rate at each redshift.

\subsection{The growth of SMBH}\label{subsec:smbhmodel}
Let us briefly summarize the growth model of SMBHs introduced by \cite{Enoki03}.
In this model, it is assumed that SMBHs grow by coalescence when their 
host galaxies merge and are fueled by accreted
 cold  gas during major mergers of galaxies. 
 When the host galaxies merge, pre-existing SMBHs
sink to the center of the new merged galaxy due to dynamical friction
(or other mechanisms such as gas dynamics), evolve to
the gravitational wave emission regime and eventually coalesce. Although
the timescale for this process is unknown, for
the sake of simplicity  we assume that SMBHs instantaneously evolve to
the gravitational wave emission regime and coalesce. 
Gas-dynamical simulations have demonstrated that the major merger of
 galaxies can
drive substantial gaseous inflows and trigger starburst activity
(e.g. \citealp{Mihos94}, \citeyear{Mihos96}; \citealp{Barnes96}).
Thus, it is reasonable to assume that 
during a major merger of galaxies,
a certain fraction of the cold gas
that is proportional to the total mass of stars 
newly formed at starburst accretes onto the newly formed SMBH
 and this accretion process leads to a quasar activity. 
Under this assumption, the mass of cold gas accreted on a SMBH is given by  
\begin{eqnarray}  
 M_{\rm acc} &=& f_{\rm BH} \Delta M_{*, \rm burst}, \label{eq:bhaccret}
\end{eqnarray} 
where $f_{\rm BH}$ is a constant and $\Delta M_{*, \rm burst} $ is the
total mass of stars formed at starburst. The free parameter of $f_{\rm BH}$ is fixed 
by matching  the observed relation between a spheroid mass and
a black hole mass  $M_{\rm BH} /M_{\rm spheroid} = 0.001 - 0.006$ (e.g. \citealp{Magorrian98}); we find that the favorable value of
$f_{\rm BH}$ is nearly $0.03$. 

\begin{figure}
\epsscale{0.4}
\plotone{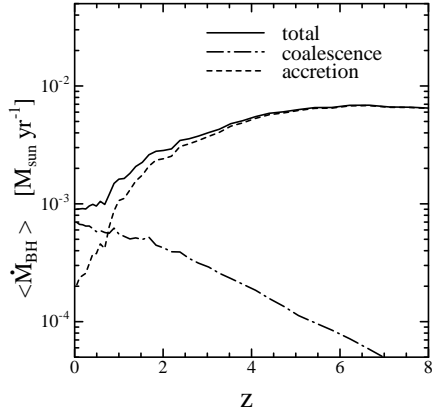}
\caption{Averaged SMBH mass growth rate, $\langle \dot{M}_{\rm BH} \rangle$, of the model. 
The solid, dot-dashed and short dashed lines indicate 
 SMBH mass growth rate of total, due to SMBH coalescence and due to gas accretion, respectively.\label{fig1}}
\end{figure}

Figure \ref{fig1} shows the model prediction of the growth rate of
the averaged SMBH mass, $\langle \dot{M}_{\rm BH} \rangle$. We define the
averaged SMBH mass growth rate as follows,
\begin{equation}
\langle \dot{M}_{\rm BH} \rangle = \frac{\int \dot{M}_{\rm BH} \phi_{\rm BH}(M_{\rm BH},z)dM_{\rm BH}}{\int \phi_{\rm BH}(M_{\rm BH},z) dM_{\rm BH} }, \label{eq:bhmassgrowth} 
\end{equation} 
where $ \dot{M}_{\rm BH}$ is the mass increase rate of SMBH with $M_{\rm
BH}$ at $z$, and $\phi_{\rm BH}(M_{\rm BH},z)$ is the black hole mass function.
$\dot{M}_{\rm BH}$ is given by $\dot{M}_{\rm BH} = (M_{\rm acc} + M_{\rm
coal})/\Delta t$ where $M_{\rm coal}$ is the mass increment due to SMBH
binaries coalescence (the mass of smaller SMBH).
This figure shows
the SMBH mass growth is mostly due to gas accretion and does not
depend on 
initial seed masses. At lower redshift ($z \lesssim 1$), since the cold gas is depleted by
star formation, the SMBH coalescence becomes a dominant process of the SMBH
mass growth. 

\begin{figure}
\epsscale{0.75}
\plotone{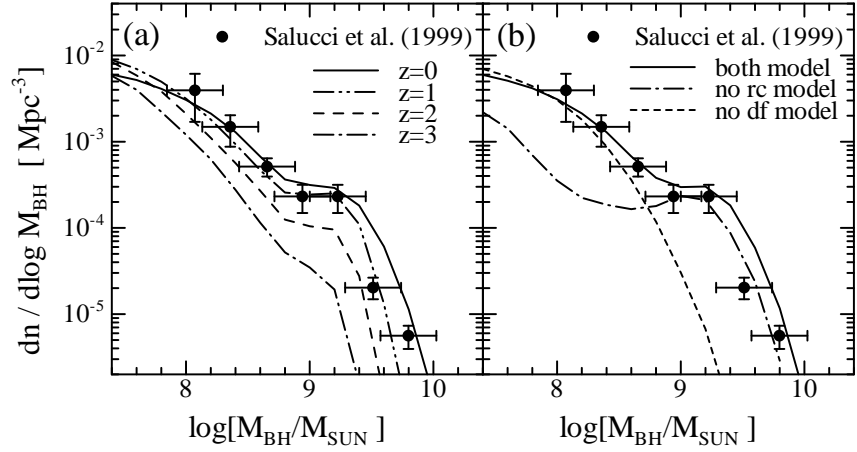}
\caption{(a) Black hole mass function of the model 
 at a series of redshifts. The solid, dot-dot-dashed, dashed and dot-dashed lines indicate
 the results at $z=0, 1, 2$, and $3$, respectively. The symbols with
 errorbars are the
 present black hole mass function obtained by \citet{Salucci99}. 
(b) Black hole mass function of models for three models at $z=0$. The
 solid, dot-dashed and short dashed lines indicate 
 both model, no random collision model (no rc model) and no
 dynamical friction model (no df model), respectively. Both model
 includes dynamical friction and random collision as the galaxy merging
 mechanism. \label{fig2}}
\end{figure}

Figure \ref{fig2} (a) shows the black hole mass functions in our
model at a series of various redshifts. This indicates that the number density
of the most massive black holes increases monotonically with time in our
scenario in which SMBHs grow by an accretion of cold gas and by
coalescence. In this figure, we superpose the black hole mass function at $z=0$
obtained by \citet{Salucci99}. 
The present mass function in our model is quite consistent with the
observation. 
Our galaxy formation model
includes dynamical friction and random collision as galaxy merging
 mechanisms. For comparison, in figure \ref{fig2} (b), we also plot
 the black hole mass functions at $z=0$ of other
two models: no random collision model (no rc model) and no
 dynamical friction model (no df model). In the no rc model and the 
no df model,
mergers owing to random collision and dynamical friction are
switched off, respectively.
 This figure shows that the mass function for low mass 
black holes are determined by random
collisions between satellite galaxies and that for high
mass black holes are influenced by dynamical friction.
The shape of black hole mass function depends on detailed gas processes.
The important contribution of mass increment of SMBHs in central galaxies is
 the cold gas, which is accreted to only central galaxies. 
SNe feedback removes this cold
gas more efficiently in smaller galaxies with $V_{\rm circ} < V_{\rm
hot} = 280 {\rm km~s}^{-1}$ ($M_{\rm gal} < 10^{12} M_{\odot}$). Thus, the growth of the SMBHs in small central
galaxies suffers from SNe feedback. In the no rc model, SMBHs
mainly exist central galaxies. Therefore, the shape of the
black hole mass function in the no rc model has a bump at high mass end
($M_{\rm BH} \sim 10^{9} M_{\odot}$) which corresponds to SMBHs in the
central galaxies ($M_{\rm gal} \sim 10^{12} M_{\odot}$). 
On the other hand, in the no df model, high mass SMBHs cannot be produced since
galaxies cannot merge with the massive central galaxy.

\cite{Enoki03} investigated environments and clustering
properties of quasars using this SA model which can 
also reproduce quasar luminosity functions
at different redshifts.
While our approach is similar to \cite{KH00}, their star formation and
feedback models are different from ours and their model does not include
the random collision process. Therefore, their resultant
model description differs from ours in equation (\ref{eq:bhaccret}).
Using the SA model incorporated with this SMBH growth model, we estimate
the comoving number density, $n_c(M_1,M_2,z) dM_1 dM_2 dz$, 
of the coalescing SMBH binaries with mass $M_1 \sim M_1 + dM_1$ and $M_2
\sim M_2 +dM_2$ at $z$ in a given redshift interval $dz$. 

It is difficult to know how SMBH binaries manage to shrink
to a gravitational wave emission regime after their host galaxies merge,
because all physical processes and conditions related to this problem
are still not clear
(dynamical friction, stellar distribution, triplet SMBH interaction, gas
dynamical effects and so on). Nevertheless, many authors have tackled this problem
\citep[e.g.][]{Begelman80,Makino97s,Gould00,Yu02,Armitage02,Blaes02,Milosavljevic03}.
In what follows, we assume that SMBHs coalesce simultaneously when host
galaxies merge for simplicity. In other words, the efficiency 
of SMBH coalescence is assumed to be maximum. 
Note that \cite{Volonteri03} constructed 
a SMBH growth model in which the merging
history of dark halos with SMBHs in their center is derived from cosmological
Monte-Carlo simulations. 
Although they incorporated dynamical evolution of SMBH binaries and
triplet SMBH interactions into their model, no galaxy
formation process is taken into account.

\section{Gravitational Radiation}\label{sec:GWR}

\subsection{Background radiation from SMBH binaries}\label{subsec:gwbg}
In order to calculate the spectrum of gravitational wave background radiation
from SMBH binaries, 
it is necessary to follow 
the orbit evolution of each binary. 
However, it is difficult to set initial conditions.
To avoid this problem, 
we adopt a formulation by \cite{Jaffe03} (see also
\citealp{Rajagopal95}). 

We start calculation of the expected 
amplitude of gravitational radiation emitted by a
binary. We assume that a binary
orbit is circular for simplicity
\footnote{Note that the initial orbit of a binary 
is not necessarily circular.
Since the amplitude and timescale of gravitational wave emissions are sensitive to 
eccentricity of a binary orbit \citep{Peters63}, 
the initial distribution of eccentricity affects the final expected
amplitudes. However, we assume that the orbit becomes circular
immediately.}.
 The amplitude is given by \citep{Thorne87} 
\begin{eqnarray}
  h_{\rm s}(z,f,M_1,M_2) &=& 4\sqrt{\frac{2}{5}} \; \frac{\left(G{M_{\rm chirp}}\right)^{5/3}}{c^4 D(z)}
  \left(2\pi f_p\right)^{2/3} \nonumber \\
&=& 3.5 \times 10^{-17} \left(\frac{M_{\rm chirp}
			}{10^8 \ M_{\odot}}\right)^{5/3} \left[\frac{D(z)}{1 \ {\rm Gpc}} \right]^{-1} \left[\frac{f (1+z)}{10^{-7} \ {\rm Hz}} \right]^{2/3}
,   \label{eq:strain}
\end{eqnarray}
where ${M_{\rm chirp}}=[M_1 M_2 (M_1 + M_2)^{-1/3}]^{3/5}$ is the chirp
mass of the system and $c$ is the speed of light. $f_p$ is  the
reciprocal of the proper rest-frame period of the binary, which is
related to the observed frequency, $f$, of the
gravitational wave from the binary in a circular orbit as $f=2f_p/(1+z)$. 
$D(z)$ is the comoving
distance to the binary
\begin{equation}
D(z) = \frac{c}{H_0} \int_{0}^{z} \frac{dz'}{\sqrt{\Omega_{\rm m}(1+z')^3+ \Omega_{\Lambda}}}.
\label{eq:comdistance}
\end{equation}

Next, we calculate $\nu(M_1,M_2,z) \;dM_1 \;dM_2 \; dz$, the number of
coalescing SMBH binaries  with mass $M_1$ and $M_2$ in mass intervals $dM_1$ and $dM_2$
per observers' time occurring at $z$ in a
given redshift interval $dz$,
\begin{eqnarray}  
\nu(M_1,M_2,z) &=&  \frac{dV}{dt_0} n_c(M_1,M_2,z)  \nonumber\\
 &=& \frac{1}{(1+z)} \frac{dV}{dt_p} n_c(M_1,M_2,z),
 \label{eq:mergernum}
\end{eqnarray} 
where $t_0$ is the observers' time, $t_p$ is the proper rest-frame time
at $z$ and 
$dV$ is the comoving volume element at $z \sim z +dz$ given by
\begin{equation}
\frac{dV}{dt_p} = 4 \pi (1+z) c D(z)^2. \label{eq:comvol}  
\end{equation}

The timescale emitting gravitational waves
of a binary measured in the rest-frame is 
\begin{eqnarray}
\tau_{\rm GW} & \equiv & f_p \frac{dt_p}{df_p}. \label{eq:defgwtimescale} 
\end{eqnarray}
Since the timescale in the observer-frame is $\tau_{\rm GW, obs} =
\tau_{\rm GW} (1+z)$, for a binary in a circular orbit at redshift $z$, 
it is expressed as \citep{Peters63}
\begin{eqnarray}
\tau_{\rm GW, obs}(M_1,M_2,z,f)  & = & \frac{5}{96} \left(\frac{c^3}{G M_{\rm chirp}} \right)^{5/3} \left[
2 \pi f_p \right]^{-8/3} (1+z) \nonumber \\
 &=& 1.2 \times 10^4 \left(\frac{M_{\rm chirp}
			}{10^8 \ M_{\odot}}\right)^{-5/3} \left[\frac{f}{10^{-7} \ {\rm Hz}} \right]^{-8/3} (1+z)^{-5/3}\ \ {\rm yr}. \label{eq:gwtimescale} 
\end{eqnarray}
As a SMBH binary evolves with time, the 
frequency becomes higher. We assume that 
the binary orbit is quasi-stationary (i.e. phase evolution
time scale is less than $\tau_{\rm GW}$) until
the radius equals to $3 R_{\rm S}$, where $R_{\rm S}$ is the Schwarzschild
radius : the radius of the innermost stable circular orbit (ISCO) for
a particle and a non-rotating black hole. 
Then the maximum  frequency $f_{\rm max}$ is 
\begin{eqnarray}
f_{\rm max}(M_1,M_2,z) &=& \frac{c^3}{6^{3/2} \pi G M_1 (1+z)} \left(1+\frac{M_2}{M_1} \right)^{1/2} \nonumber \\
&=& 4.4 \times 10^{-5} (1+z)^{-1} \left(\frac{M_1}{10^8 M_{\odot}}\right)^{-1} \left(
1+\frac{M_2}{M_1} \right)^{1/2} {\rm Hz}, \label{eq:fmax}
\end{eqnarray}
where $M_1$ and $M_2$ are SMBH masses ($M_1 > M_2$).
The number of coalescencing SMBH binaries as a source of gravitational wave
source with observed frequency $f$ at redshift $z$
is estimated as
\begin{equation} 
\nu(M_1,M_2,z) dz \; dM_1 \; dM_2 \; \tau_{\rm GW, obs} \; \theta (f_{\rm max} - f), 
\end{equation}
where $\theta(x)$ is the step function. Therefore, 
from equations (\ref{eq:strain}), (\ref{eq:mergernum}),
(\ref{eq:gwtimescale}) and (\ref{eq:gwspec}), we finally obtain
the spectrum of the gravitational wave background
radiation, 
\begin{eqnarray}
  h^2_c(f) & = & \int dz \; dM_1 \; dM_2 \; h_{\rm s}^2 \; \nu(M_1,M_2,z) \; \tau_{\rm GW, obs} \; \theta (f_{\rm max} - f).    \label{eq:gwspec}
\nonumber
\\
& = & \int dz \; dM_1 \; dM_2 \; \frac{4 \pi c^3}{3}
 \left(\frac{G M_{\rm chirp}}{c^3} \right)^{5/3} (\pi f )^{-4/3} (1+z)^{-1/3}
 n_c(M_1,M_2,z) \; \theta (f_{\rm max} - f). 
\nonumber
\\
\label{eq:spectrum}
\end{eqnarray}
We note that the equation (\ref{eq:spectrum}) agrees 
with the result of \cite{Phinney01} except for 
the step function, $\theta (f_{\rm max} - f)$. The original formulation
of \cite{Jaffe03} does not include this effect, either. 

\begin{figure}
\epsscale{0.75}
\plotone{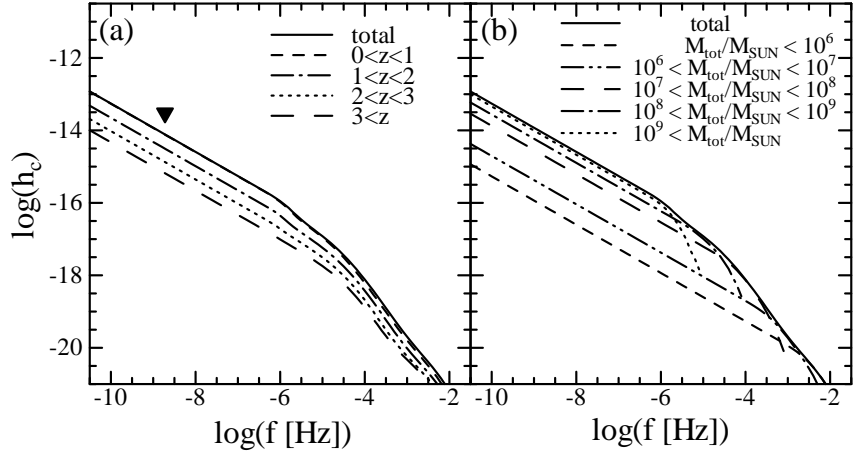}
\caption{Spectrum of gravitational wave background radiation, $h_{\rm c} (f)$, from SMBH binaries in different redshift intervals (a) and in different total mass ranges(b). 
(a) The total spectrum (solid line) and the other lines show those
 in different redshift intervals $0 \le z<1$ (dashed line), $1\le z<2$
 (dot-dashed), $2\le z<3$ (short dashed) and $3 \le z$ (long dashed). The filled reverse triangle
 shows the current limit from pulsar timing measurements \cite{Lommen02}.
(b) The total spectrum (solid line) and the other lines show those
 in different total mass intervals $M_{\rm tot} \le 10^6 M_{\odot}$
 (dashed line), $10^6 M_{\odot} < M_{\rm tot} \le 10^7 M_{\odot}$
 (dot-dot-dashed),
 $10^7 M_{\odot} < M_{\rm tot} \le 10^8 M_{\odot}$ (long dashed),
$10^8 M_{\odot} < M_{\rm tot} \le 10^9 M_{\odot}$ (dot-dashed)
 and $10^9 M_{\odot} < M_{\rm tot} $ (short dashed).
 \label{fig3}}
\end{figure}

As shown in figure \ref{fig3}, 
the spectrum changes its slope at $f \sim 1 \mu {\rm Hz}$
owing to lack of power associated with the upper limit frequency,
$f_{\rm max}$. This feature is consistent with the results of
\cite{Wyithe03} and \cite {Sesana04}.
The predicted  strain spectrum is $h_{\rm c}(f) \sim 10^{-16} (f/1 \mu
{\rm Hz})^{-2/3}$ for $f \lesssim 1 \mu {\rm Hz}$, just
below the current limit from the pulsar timing measurements by
\cite{Lommen02}. 
In our model, we assume that SMBHs coalesce simultaneously when their host
galaxies merge. Therefore, the efficiency of SMBH coalescence is maximum
and the predicted amplitude of gravitational wave spectrum should be
interpreted as the upper limit. 

\begin{figure}
\epsscale{0.42}
\plotone{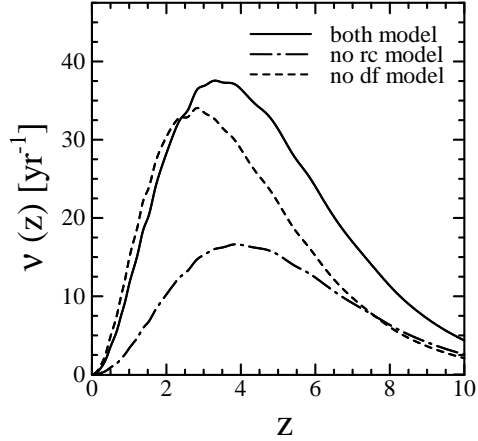}
\caption{SMBH coalescence rate in observers' time unit a year, $\nu(z)$. 
The solid, dot-dashed and short dashed lines indicate 
 both model, no random collision model (no rc model) and no
 dynamical friction model (no df model), respectively. Both model
 includes dynamical friction and random collision as the galaxy merging
 mechanism.\label{fig4}}
\end{figure}

In figure \ref{fig3} (a), we plot the spectra from binaries in
different redshift intervals. This figure shows that the total spectrum
of background radiation comes from coalescing SMBH binaries 
at low redshift, $0 \le z<1$. 
In figure \ref{fig4}, we plot the SMBH
coalescence rate in observers' time unit a year, $\nu(z) = \int \nu(M_1,M_2,z)
\;dM_1 \;dM_2 $. This figure indicates that  
the coalescence rate at low redshift is lower than the
coalescence rate at high redshift. However, 
the main contribution to the  background radiation is gravitational wave
from the coalescing SMBH binaries at low redshift, $0 \le z<1$, because 
the distance from an observer to the SMBH binaries at low redshift is
shorter and the mass of SMBHs at low redshift is higher.
In figure \ref{fig3}(b), we also plot the spectra from binaries in different
 total mass intervals ($M_{\rm tot} = M_1+M_2$). This figure shows that for $f \gtrsim
 10^{-4} {\rm Hz}$ the total
 spectrum of background radiation comes from coalescing SMBH binaries
 with total mass $M_{\rm tot} \lesssim 10^8 M_{\odot}$. 

\begin{figure}
\epsscale{0.75}
\plotone{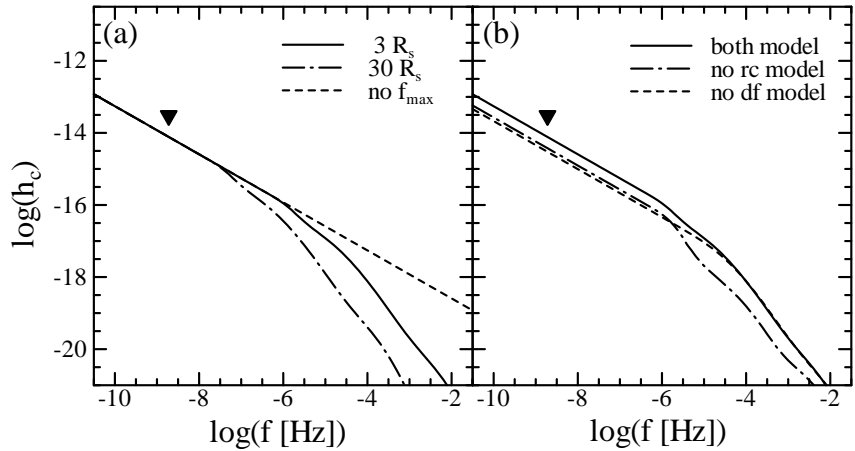}
\caption{Spectrum of gravitational wave background radiation, $h_{\rm c} (f)$,
in different $f_{\rm max}$ (a) and in different galaxy merger models (b). 
 The filled reverse triangle
 shows the current limit from pulsar timing measurements \cite{Lommen02}.
(a) The solid, dot-dashed and short dashed lines indicate the results
 with $f_{\rm max}$ corresponding to $3R_{\rm S} $, $30 R_{\rm S}$
and no frequency cut off, respectively.
(b)  The solid, dot-dashed and short dashed lines indicate 
 both, no random collision model (no rc model) and no
 dynamical friction model (no df model), respectively. Both model
 includes dynamical friction and random collision as the galaxy merging
 mechanism. \label{fig5}}
\end{figure}

When SMBHs are spinning and/or when their
masses are comparable, the definition of ISCO becomes vague and
our assumption that the cutoff frequency $f_{\rm max}$ corresponds to 
$3 R_{\rm S}$ may not be correct. 
To see the effects of the cutoff frequency, we plot the spectra for
different values of $f_{\rm
max}$, corresponding to $3  R_{\rm S}$, $30 R_{\rm S}$, and no frequency
cut off, respectively (figure \ref{fig5} (a)). Lowering $f_{\rm
max}$ causes a suppression in the stochastic 
background at high frequencies $f\lesssim 10^{-7}$Hz 
since a large portion of high frequency modes are cut off. As shown in
this figure, the cut off radius does not affect the amplitude at the
frequency of the pulsar timing measurement $f \simeq 10^{-9}{\rm Hz}
$. 

Our galaxy formation model incorporates dynamical friction and random
collision as galaxy merging mechanisms. In order to examine the effect 
of these two galaxy merger mechanisms on the spectrum of gravitational 
wave background radiation, in figure \ref{fig5} (b), we also plot the 
spectrum of background radiation of other two models: no rc model and no
df model. 
 The no df model can not produce higher mass SMBH (see
 fig. \ref{fig2}(b)). Consequently, the spectrum in no df model bends
 at larger frequency since the number of SMBHs with smaller  $f_{\rm max}$
  decreases. 
Furthermore, in the no df model,
 the amplitude of gravitational waves from each binary becomes smaller
. 
However, the coalescence rate in no df model is 
 higher than the rate in no rc model as shown in figure \ref{fig4}. 
As a result, the amplitude of the spectrum in no df model is
 roughly equal to the amplitude in no rc model.

\subsection{Gravitational wave burst from SMBH coalescence}\label{subsec:burst}
After an inspiraling phase, SMBHs plunge into a final death inspiral 
 and merge to form a single black hole. We call a set of a 
plunge and a subsequent 
early non-linear ring-down phase as a {\it burst}\footnote{Later ring-down
phase at linear perturbation regime is not included. }.
In this subsection, we estimate 
the expected burst event rate per observers' time
using the amplitude of burst gravitational wave given by  \cite{Thorne76}, 
and the SMBH coalescence rate calculated by our SA model, $n_c(M_1,M_2,z)$. 

The gravitational wave amplitude is given by
\begin{equation}
h^{2} = \frac{2 G F_{\rm GW}}{\pi c^3 f_{\rm c}^2}, \label{eq:burstamp}
\end{equation}
where $F_{\rm GW}$ is the energy flux of the gravitational wave at the
observer and $f_{\rm c}$ is the observed characteristic frequency.
$f_{\rm c}$ from the gravitational
wave burst occurring at $z$ is
\begin{eqnarray}
f_{\rm c} &=&  \frac{c^3}{3^{3/2} GM_{\rm tot} \ (1+z)} \nonumber \\
 &=& 3.9 \times 10^{-4} \left(\frac{M_{\rm tot}}{10^8 \ M_{\odot}}\right)^{-1}
 (1+z)^{-1} {\rm Hz}, \label{eq:burstfreq}
\end{eqnarray}
where $M_{\rm tot}$ is total mass of the black holes.
 The energy flux from the burst gravitational wave occurring at $z$ is given by  
\begin{eqnarray}
F_{\rm GW} &=& \frac{\epsilon M_{\rm tot} c^2 f_{\rm c}}{4 \pi D(z)^2 (1+z)}, \label{eq:burstflux}
\end{eqnarray}
where $\epsilon$ is the efficiency of the energy release and $D(z)$ is
the distance to the source from the observer given by equation
(\ref{eq:comdistance}). From equations (\ref{eq:burstamp}),
(\ref{eq:burstfreq}) and (\ref{eq:burstflux}), the characteristic
amplitude of the burst gravitational wave is
\begin{eqnarray}
h_{\rm burst} &=& \frac{3^{3/4} \epsilon^{1/2} G M_{\rm tot}}{2^{1/2} \pi c^2 D(z)} \nonumber \\
 &=& 7.8 \times 10^{-16} \left(\frac{\epsilon }{0.1} \right)^{1/2} \left(\frac{M_{\rm tot}}{10^8 \ M_{\odot}}\right) \left[\frac{D(z)}{1 \ {\rm Gpc}} \right]^{-1}. \label{eq:ampburst} 
\end{eqnarray}
Equations (\ref{eq:burstfreq}) and (\ref{eq:ampburst}) show that we can
estimate the amplitude and the characteristic frequency of the burst
gravitational wave, once we know the total mass and redshift of
coalescing SMBHs.
 From eq.(\ref{eq:burstfreq}), eq.(\ref{eq:ampburst}) and  $ n_c(M_1,M_2,z)$,
we obtain $n_{\rm burst} (h_{\rm burst},f_{\rm c}, z) \;dh_{\rm
burst} \;df_{\rm c} \;dz $ , which is the comoving  number density of
gravitational wave burst events occurring at $z$ in a given redshift
interval $dz$ with amplitude $h_{\rm burst} \sim h_{\rm burst} + dh_{\rm
burst}$ and with
 characteristic frequency  $f_{\rm c} \sim f_{\rm c} + df_{\rm c}$.
Then, the expected event rates of gravitational wave bursts per observers' time with amplitude $h_{\rm burst} \sim h_{\rm burst} + dh_{\rm burst}$ and
 characteristic frequency  $f_{\rm c} \sim f_{\rm c} + df_{\rm c}$,
 $\nu_{\rm burst}(h_{\rm burst},f_{\rm c})  \;dh_{\rm burst} \;df_{\rm
 c} $ is given by  
\begin{eqnarray}
\nu_{\rm burst}(h_{\rm burst},f_{\rm c}) &=& \int n_{\rm burst} (h_{\rm burst},f_{\rm c}, z) \frac{dV}{dt_0} dz. \label{eq:burstrate}
\end{eqnarray}
The integrated event rates of gravitational wave bursts per
 observers' time with amplitude $h_{\rm burst} \sim h_{\rm burst} + dh_{\rm
 burst}$, $\nu_{\rm burst}(h_{\rm burst})  \;dh_{\rm burst}$, 
are given respectively as follows,
\begin{equation}
\nu_{\rm burst}(h_{\rm burst}) = \int \nu_{\rm burst}(h_{\rm burst},f_{\rm c})
d f_{\rm c}. \label{eq:hburstrate}
\end{equation}

\begin{figure}
\epsscale{0.9}
\plottwo{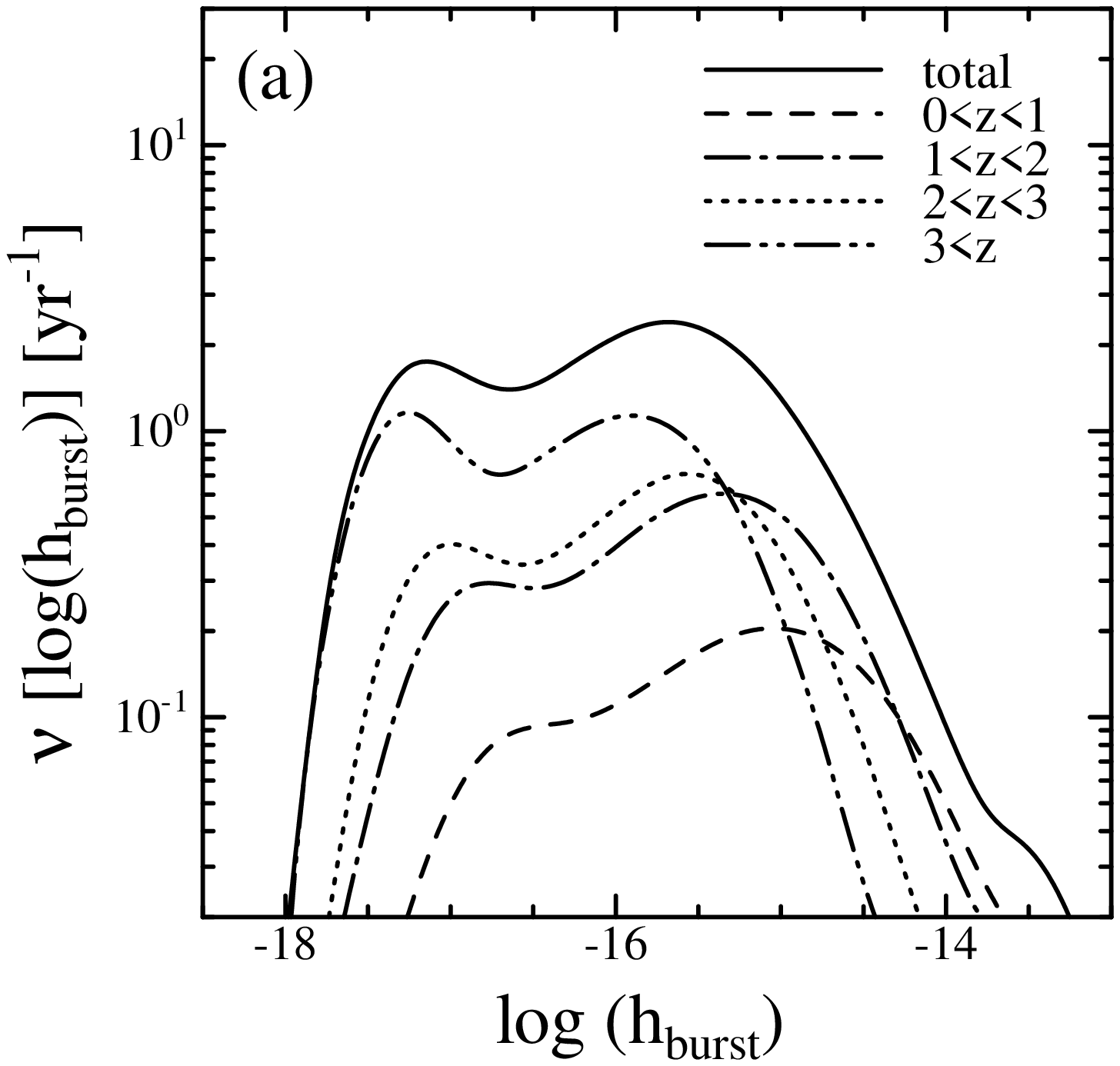}{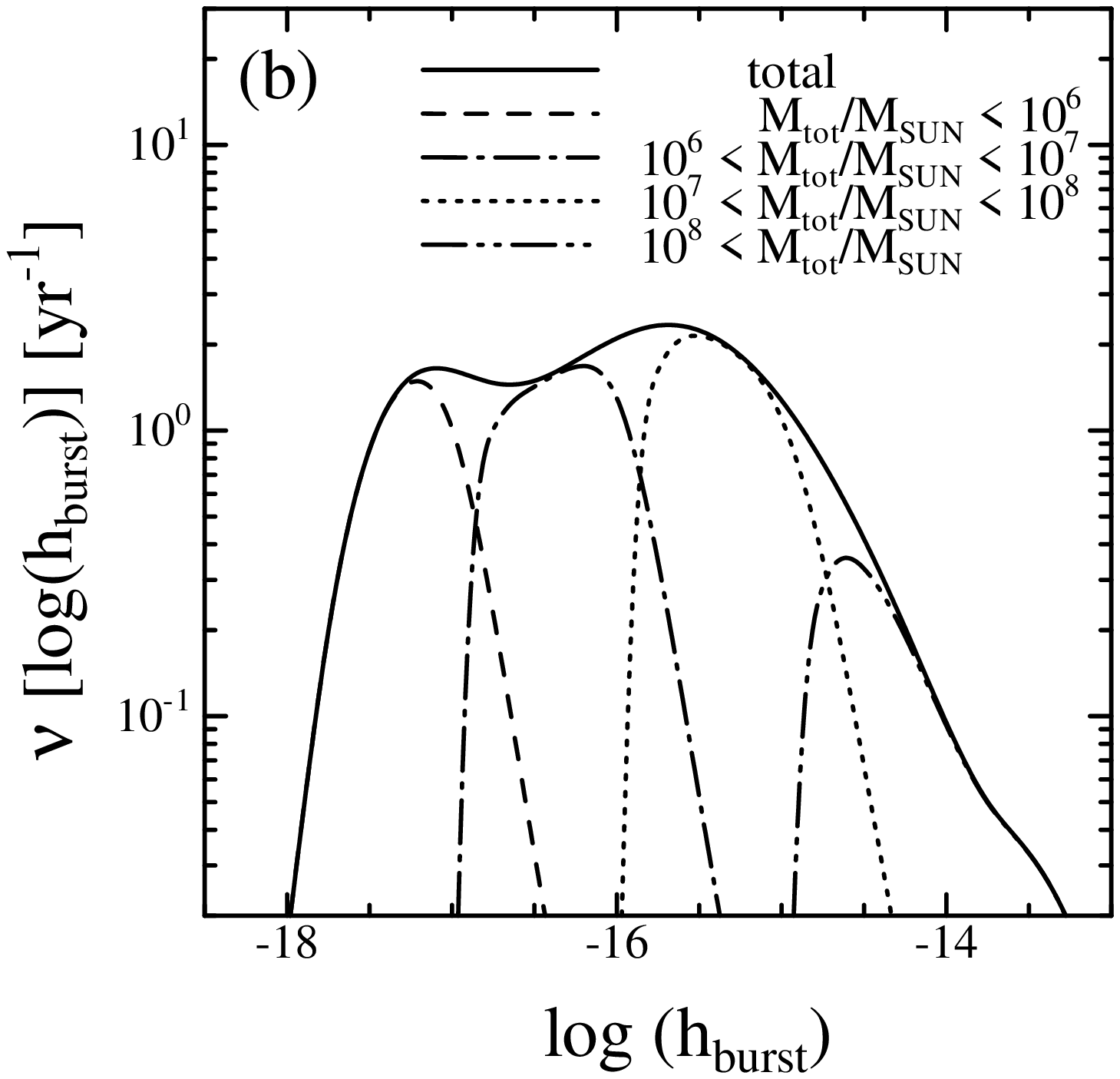}
\caption{Integrated event rate of gravitational wave burst per
 observers' time  unit a year $\nu_{\rm burst} [\log(h_{\rm burst})]$. (a) The total integrated event
 rate  (solid line) and the other lines show those
 in different redshift intervals $0\le z<1$ (dot), $1\le z<2$
 (dot-dashed), $2 \le z<3$ (short dashed) and $3 \le z$ (dot-dot-dashed).
(b) The total integrated event
 rate  (solid line) and the other lines show those
 in different black hole mass  intervals $M_{\rm tot} \le 10^{6} M_{\odot}$ (dashed line), $ 10^{6} M_{\odot} < M_{\rm tot} \le 10^{7} M_{\odot}$
 (dot-dashed), $ 10^{7} M_{\odot} < M_{\rm tot} \le 10^{8} M_{\odot}$ (dot) and $ 10^{8} M_{\odot} < M_{\rm tot} $ (dot-dot-dashed).
\label{fig6}}
\end{figure}

In figure \ref{fig6}, we plot the total integrated event rates of
gravitational wave bursts and integrated event rates in
different redshift intervals (figure \ref{fig6}(a)) and in different total mass intervals in (figure \ref{fig6}(b)).
Here we set the efficiency of the energy release
$\epsilon = 0.1$, while the precise value of this 
parameter is unknown. 
\cite{Flanagan98} argued that the efficiency
could reach 10 \%, which depends on alignments of spin and
parameter choices. In most typical events, a conversion efficiency will
 probably be a few percent \citep{Baker-etal01}. 
From equation (\ref{eq:ampburst}), 
one can see that the change of efficiency results in the 
parallel displacement in the horizontal direction
in figure \ref{fig6}. 
The shape of $\nu_{\rm burst}(h_{\rm burst})$ reflects the black hole
mass functions and the SMBH coalescence rates, which depend on the
complex galaxy formation processes.
In figure \ref{fig6}  (a), one can notice that 
there are two peaks in the event rate in terms of $h_{\rm burst}$.  
A peak at $h_{\rm burst} \sim 10^{-17}$ corresponds to bursts 
from SMBH binaries with $M_{\rm
tot} < 10^6 M_{\odot}$ whose total number
is the largest at high redshift $z>3$.
Another peak at $h_{\rm burst} \sim 10^{-15}$
corresponds to bursts from SMBH 
binaries with  $ 10^7M_{\odot} < M_{\rm tot} < 10^8 M_{\odot}$ 
whose coalescence probability is the largest at high redshift $z>3$.
Figure \ref{fig6} (b) indicates that burst signals with large amplitude 
($h_{\rm burst} \gtrsim 10^{-15}$) correspond to coalescence of
``massive'' SMBH binaries   
with $M_{\rm tot} \gtrsim 10^8 M_{\odot}$ occurring at $z
\lesssim 1$. This is because the distance from the earth 
to SMBHs at low redshift is
shorter and the mass of SMBHs at low redshift is larger. 
On the other hand,  burst signals with amplitude 
($h_{\rm burst} \gtrsim 10^{-15}$) corresponds to
coalescence events of 
``less massive''  SMBH
binaries with $M_{\rm tot} \lesssim 10^7 M_{\odot}$ 
occurring at $z \gtrsim 2$. These events dominate the 
expected burst event rate provided that the sensitivity 
of the detector is sufficiently good.
This feature is quite important because it breaks the degeneracy
between mass and distance.
Our model predicts that the expected rates of the coalescence events owing to  
SMBH binaries with small mass $M_{\rm tot} < 10^6 M_{\odot}$ at low redshift 
and those with large mass $M_{\rm tot} > 10^8 M_{\odot}$ at high redshift 
$z>3$ are small. 

\begin{figure}
\epsscale{0.8}
\plotone{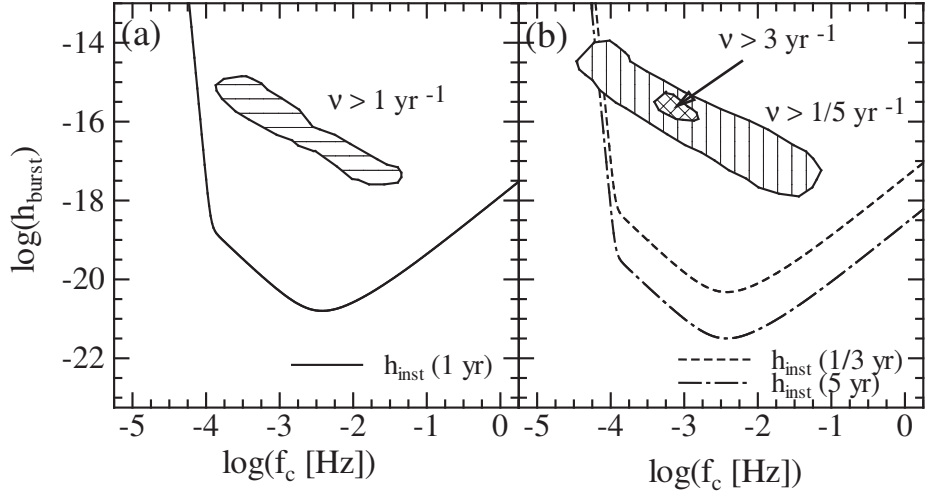}
\caption{Expected signals of gravitational burst from SMBH. (a)
The horizontally hatched area shows
 the region, $\nu_{\rm burst} [\log(h_{\rm burst}),\log(f_{\rm c})] > 1 {\rm yr}^{-1}$. The solid curve indicates the instrumental noise threshold for one year of {\sl LISA} observations. (b) The vertically hatched area shows the region, $\nu_{\rm burst} [\log(h_{\rm burst}),\log(f_{\rm c})] > 1/5 {\rm yr}^{-1}$ and the diagnal cross-hatched area show the region, $\nu_{\rm burst} [\log(h_{\rm burst}),\log(f_{\rm c})] > 3 {\rm yr}^{-1}$. The dot-dashed and the short dashed lines indicates the instrumental noise threshold for $5$ year  and $1/3$ year of {\sl LISA} observations, respectively. 
 \label{fig7}}
\end{figure}

Figure \ref{fig7} shows that the expected region for signal of
gravitational wave bursts and 
the instrumental noise threshold for {\sl LISA}, $h_{\rm
inst}$. For randomly oriented sources, a sensitivity for a search of 
gravitational wave bursts in an observation time $T_{\rm obs}$ is given by \citep{Thorne87,Haehnelt94} 
\begin{equation}
h_{\rm inst}^{2} \sim \left(\frac{T_{\rm obs}}{1 {\rm yr}}\right)^{-1} 10 S_{\rm h} f_{\rm c},
\end{equation}      
where $S_{\rm h}$ is the spectral instrumental noise density. 
We  compute $h_{\rm inst}$
from the fitting formula for $S_{\rm h}$ of {\sl LISA} \citep{Hughes02}. The expected region for $\nu_{\rm burst} [\log(h_{\rm
burst}),\log(f_{\rm c})] > 1 \ {\rm yr}^{-1}$ is above this instrumental
noise threshold. For comparison, we show the region for  $\nu_{\rm burst} [\log(h_{\rm
burst}),\log(f_{\rm c})] > 1/5 \ {\rm yr}^{-1}$ and  $\nu_{\rm burst} [\log(h_{\rm
burst}),\log(f_{\rm c})] > 3 \ {\rm yr}^{-1}$
 in figure \ref{fig7} (b).

From figure \ref{fig6} and \ref{fig7}, we conclude that the
{\sl LISA} can 
detect intense bursts of gravitational waves at a rate of  
 $\sim 1.0  {\rm yr}^{-1}$ assuming
that dominant part of these burst events occur at
$z \gtrsim 2$. 
Even in the case of $\epsilon = 0.001$, the
{\sl LISA} can detect intense bursts of gravitational
waves in one year observation, since 
$h_{\rm burst} \propto \epsilon^{1/2} $. 
In addition, we find that large
amplitude $h_{\rm burst} \sim 10^{-15}$
signals correspond to coalescence events of massive 
SMBH binaries $M_{\rm tot} \sim 10^8
M_{\odot}$
at low redshift $ z \lesssim 1$ and small amplitude $h_{\rm
 burst} \sim 10^{-17}$  signals correspond to less massive 
SMBH binaries $M_{\rm tot} \sim 10^6
M_{\odot}$ at high redshift $ z \gtrsim 3$. 

Based on a SA-model \citep{KH00}, 
\cite{Haehnelt03} concluded that {\sl LISA} 
might detect SMBH coalescence events at a rate 
$0.1 \sim 1.0 {\rm ~yr}^{-1}$ over the redshift range of $0 \le z \le 5$
although no explicit calculation on gravitational wave emission is
done. Our result is consistent with his result.

\section{Summary and Conclusions}\label{sec:summary}

In this paper, we have estimated the coalescence rate of supermassive
black hole (SMBH) binaries in the
centers of galaxies using a new semi-analytic model of
galaxy and quasar formation (SA model) given by \cite{Enoki03} based on the hierarchical structure formation
scenario. Then, 
we calculated the spectrum of the gravitational wave
background from inspiraling SMBH binaries based on the formulation given
by \cite{Jaffe03} and estimated the expected amplitudes and 
event rates of intense bursts of
gravitational waves from coalescing SMBH binaries.

Our SA model includes dynamical friction and random collision as galaxy merging mechanisms, and assumes that a SMBH is fueled by accretion of cold gas
during a  major merger of galaxies leading to a spheroid formation,
and that SMBHs coalesce simultaneously when host galaxies
merge. Many previous other studies have paid attention to only SMBH growth
and did not take galaxy formation processes into account. For
investigating the relations between SMBH growth and galaxy formation
processes, SA methods of galaxy and SMBH formation are suitable tools 
\citep[]{KH00,Cattaneo01,Enoki03, Menci03,Granato04}.   
Our SA model can reproduce not only observational properties of galaxies, 
but also the present SMBH mass function and the quasar luminosity functions
at different redshifts \citep{Nagashima01b,Enoki03}.

We have found that the gravitational wave background radiation spectrum for
$f \lesssim 1 \mu {\rm Hz}$ has a characteristic
strain $h_c(f) \sim 10^{-16} (f/1 \mu {\rm Hz})^{-2/3}$ just below the
 detection limit from the current measurements of the pulsar timing.
The slope of the spectrum for $f \gtrsim 1 \mu {\rm Hz}$ gets steep
owing to the upper limit in frequency set by 
the radius of the innermost stable circular orbit. 
The stochastic 
background radiation mainly comes from
inspiraling SMBH binaries at $0<z<1$. Therefore, the background radiation
can probe inspiraling SMBH binaries at low redshift.

We have also found that {\sl LISA} might detect
 intense bursts of gravitational waves owing to the SMBH coalescence events
at a rate $0.1 \sim 1.0 {\rm ~yr}^{-1}$ and that the main
contribution to the event rate comes from 
SMBH binary coalescence at high redshift $z \gtrsim 2$.
Our model predicts that burst signals with a large amplitude 
correspond to coalescence of large mass SMBH  binaries 
at low redshift while those with a small amplitude 
correspond to coalescence of small mass SMBH binaries at high redshift. 
This prediction can be tested by future measurements 
of the amplitude and the phase evolution in 
gravitational waves from inspiraling SMBH binaries \citep{Hughes02}. 
Comparing these
predictions with observations in future, 
we can put a stringent 
constraint on SMBH formation and evolution models.

\acknowledgments
We thank  N. Gouda, K. Okoshi, S. Yoshioka and H. Yahagi for useful
comments and discussions. We also thank K.S. Thorne and S. Hughes for 
useful information. MN acknowledges Research Fellowships of the Japan Society
for the Promotion of Science for Young Scientists (No.00207) and support from the PPARC rolling grant for extragalactic
astronomy and cosmology at Durham.
NS is supported by Japanese Grant-in-Aid for Science Research Fund of
the Ministry of Education, No.14340290.
Numerical computations in this work were partly carried out at the
Astronomical Data Analysis Center of the National Astronomical
Observatory Japan.


\clearpage

\begin{table}
\caption{Model parameters.}  
\begin{center}
\label{table1}
\begin{tabular}{ccccccccccccc}
\hline
\hline
 \multicolumn{5}{c}{Cosmological parameters} &
& \multicolumn{6}{c}{Astrophysical parameters} \\
\cline{1-5} \cline{7-12}
$\Omega_{\rm m}$ & $\Omega_{\Lambda}$ &$h$ &$\sigma_{8}$ &$\Omega_{\rm b}$ & &
$V_{\rm hot}$ (km~s$^{-1}$)
& $\alpha_{\rm hot}$ & $\tau_{*}^{0}$ (Gyr)
& $\alpha_*$ & $f_{\rm bulge}$ & $\Upsilon$\\
\hline
0.3&0.7&0.7&0.9&$0.02 h^{-2}$ && 280 & 2.5 & 1.5 & -2   & 0.5 & 1.7\\
\hline
\end{tabular}
\end{center}
\end{table}


\end{document}